\documentclass[%
 reprint,
bibnotes,
amsmath,
amssymb,
 aps,
floatfix,
]{revtex4-2}

\usepackage[T1]{fontenc}

\usepackage[greek,english]{babel}

\usepackage{graphicx}
\graphicspath{{.}}

\usepackage[utf8]{inputenc}
\usepackage[T1]{fontenc}
\usepackage{xcolor}
\usepackage[hidelinks,colorlinks=false]{hyperref}
\usepackage{fullpage}
\usepackage{enumitem}
\usepackage{physics}\usepackage[right]{lineno}
\usepackage{array}
\usepackage{hhline}
\usepackage{amssymb}
\usepackage{mathtools}
\usepackage{amsmath} 
\usepackage{graphicx}
\graphicspath{{./images}}
\usepackage{subfigure}
\usepackage{nameref}
\usepackage{times}

\usepackage[mediumspace,mediumqspace,squaren]{SIunits}
\usepackage{natbib}

\usepackage[capitalise,nameinlink,sort&compress]{cleveref}

\DeclareMathOperator{\gd}{gd}

\newcommand{\matr}[1]{\mathbf{#1}}
\newcommand{\E}{{\text{e}}}
\newcommand{\I}{{\text{i}}}
\renewcommand{\vec}[1]{\mathbf{#1}}

\newcommand{\pp}[1]{({#1})}
\newcommand{\bb}[1]{\left[ #1 \right]}
\newcommand{\eqdef}{\vcentcolon =}
\newcommand{\pdiff}[2]{\frac{\partial{#1}}{\partial{#2}}}

\newcommand{\linepdiff}[2]{{\partial{#1}}/{\partial{#2}}}

\newcommand\restr[2]{{
  \left.\kern-\nulldelimiterspace
  #1 
  \vphantom{\big|} 
  \right|_{#2} 
  }}

\begin{document}

\preprint{APS/123-QED}

% \title{Morphodynamics of Plant Shoots in Changing Gravity}

%\title{Plant Growth In Changing Gravity}

\title{Active shape control by plants in dynamic environments}
%AG:(not quite happy with it)
\author{Hadrien Oliveri}
\author{Derek E. Moulton}
\author{Heather A. Harrington}
\author{Alain Goriely}
\email{goriely@maths.ox.ac.uk}
 
 \affiliation{Mathematical Institute, University of Oxford, Oxford, OX2 6GG, United Kingdom}
 
\date{\today}

 \begin{abstract}
 Plants are a paradigm for active shape control in response to stimuli. For instance, it is well-known that a tilted plant will eventually straighten vertically, demonstrating the influence of both an external stimulus,  gravity, and an internal stimulus, proprioception. These effects can be modulated when a potted plant is additionally rotated along the plant's axis, as in a  rotating clinostat, leading to intricate shapes. We use a morphoelastic model for the response of growing plants to study the joint effect of both stimuli at all rotation speeds. In the absence of rotation, we identify a universal planar shape towards which all shoots eventually converge. With rotation, we demonstrate the existence of a stable family of three-dimensional dynamic equilibria where the plant axis is fixed in space. Further, the effect of axial growth is to induce  steady behaviors, such as solitary waves. Overall, this study offers new insight into the complex out-of-equilibrium dynamics of a plant in three dimensions and further establishes that internal stimuli in active materials are key for robust shape control.
 
 %Using active morphoelastic rod theory, we explore the interaction between gravitropism (response to gravity) and proprioception. We revisit the historical clinostat experiment, studying the shape changes in a plant shoot subjected to periodic shifts in perceived gravity through slow rolling. We identify stable configurations where the plant's centerline movement halts, persisting even as the plant grows in the form of solitary waves. This study sheds light on the intricate dynamics of plants in three dimensions, exposed to varying environmental stimuli.
 % Plants move in response to their environment; in particular, shoots exhibit \textit{gravitropism} and bend to align with gravity. Revisiting the classic \textit{clinostat} experiment, we investigate the shape dynamics of a shoot subject to periodic changes in the perceived gravity stimulus imparted through slow horizontal rolling of the plant. Theoretical analysis based on the theory of active morphoelastic rods reveals the existence of stable dynamic equilibrium where the plant axis is fixed. Allowing the plant to extend in length again yields steady behaviors, such as solitons. Overall, this study offers new insight into the complex out-of-equilibrium dynamics of a plant in three dimensions, subject to changing environmental stimuli. \todo{Bit flat}
 \end{abstract}

\maketitle

 % \linenumbers

\begin{figure}[hb!]
    \centering    \includegraphics[width=.85\linewidth]{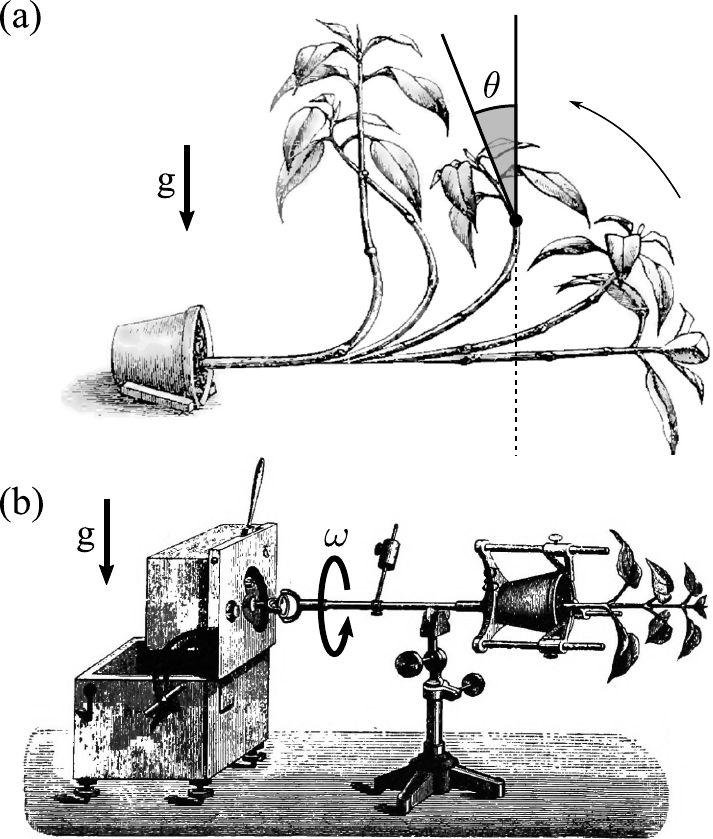}
    \caption{(a) A potted plant  realigns  itself  with  gravity when tilted horizontally. (b) In a clinostat, the effect of gravity is nullified at sufficient angular speed. In both cases, the plant's axis lies in a plane. (Adapted from \citep{pfeffer1904pflanzenphysiologie})}
    \label{fig:fig1}
\end{figure}

Active materials are characterized by their ability to adapt to external stimuli, often manifested by changes in shape. A paradigm of this adaptability is observed in the growth patterns of plant shoots, which exhibit remarkable sensitivity not only to their environment (e.g. light, gravity, wind) \cite{moulton2020multiscale} but also, intriguingly, to their own evolving shapes, a phenomenon called \textit{proprioception} \citep{moulia2019posture,hamant_how_2016}. We show that this synergistic response to multiple stimuli serves as a robust mechanism for plants to maintain structural integrity in highly dynamic environments. An important type of response in plant shoots is 
\textit{gravitropism} [\cref{fig:fig1}(a)], the tendency to react and orient their growth against the direction of gravity  \citep{moulia2009power}. While modifying gravity  experimentally is challenging, it is possible to nullify its influence by rotating the plant sufficiently fast in a \textit{clinostat} \citep{johnsson1971aspects}, shown in \cref{fig:fig1}(b).  This device, % invented and
patented by Julius von Sachs circa 1880 \citep{sachs1882vorlesungen,sachs1882orthotrope}, imparts a constant rotational motion to the plant, thereby cyclically altering the relative direction of gravity. To simulate weightlessness, the clinostat must rotate at a relatively high angular speed $\omega$, compared to the response of the plant, allowing for the averaging out of gravity's influence over multiple rotations \citep{COOK1969353}. In such a case, the plant grows straight. Further,  the general observation that growing shoots tend to straighten in the absence of other influences, indicates another well-established necessary response, called \textit{autotropism}, the  tendency to minimize curvature during growth \citep{bastien2013}.
% \todo{Not convinced (in principle, the system is invariant by rotation). A stronger argument is the work of Bastien + the observation that the sine law has no fixed point.}. %Since gravity is nullified at high angular speed, the relative influence of autotropism and gravitropism can be gauged by varying the angular speed, leading to the possibility of complex three-dimensional shapes that we study here 
Under slower rotations, the relative influence of autotropism and gravitropism can be gauged by varying the angular speed, leading to the possibility of complex three-dimensional shapes that we study here.

%In particular, numerous plants exhibit \textit{negative} gravitropism \citep[\textit{apogeotropism} in Darwin's parlance, Ref.][]{darwin1880power}, namely, their shoot bends upward, away from the Earth [\cref{fig:fig1}(a)]. 
The first model for the gravitropic response of slender shoots was formulated by Sachs in 1879 \citep{sachs1882orthotrope}. The \textit{sine law} states that the rate of change of curvature at a point is given by the sine of the inclination angle  $\theta\pp{s,t}$ between the tangent to the shoot centerline and the vertical direction, where $s$ is the arclength from the base and $t$ is the time [\cref{fig:fig1}(a)]. Recalling that the curvature is the arclength derivative of this angle, the sine law can be expressed as
\begin{equation}
\label{eqn:sachs}
    \dot\theta' + \alpha \sin \theta=0,
\end{equation}
with $\alpha$ a rate constant; and where $(\ )'$ and $\dot{(\  )}$ denote differentiation w.r.t. $s$ and $t$, respectively. Notably, unbeknownst to Sachs and his successors, the sine law is an instance of the celebrated sine-Gordon equation, a fully integrable system with a conservative structure \citep{polyanin2003handbook}; in fact, the sine law is the earliest appearance of this equation as a physical model. % with Hamiltonian density $\mathcal H  =  \theta'^2 + \dot\theta^2  - \alpha \pp{1 - \cos\theta}  $}
 While the sine law is the starting point of many augmented models \citep{bastien2013,bastien2014unifying,bastien2015unified,chelakkot2017growth,agostinelli2020nutations,moulia2022shaping}, it is restricted to planar motion and does not include autotropism, which is necessary for shoots to eventually straighten \citep{bastien2013,dumais2013beyond}. %\todo{not happy with that sentence and transition} % or the dynamic changes induced by a clinostat.

Here, we follow the plant tropism framework developed in \cite{moulton2020multiscale} to model the clinostatting plant in three dimensions as an unshearable and inextensible \textit{morphoelastic rod} \citep{moulton2013,Goriely2017} of length $\ell$. We neglect self-weight and centrifugal effects, which is valid for small shoots and slow rotation (i.e. $ \rho  g \ell^3 \ll B$ and $\rho\omega^2  \ell^4 \ll B$, with $B$ and $\rho$ denoting the bending stiffness and the linear density, respectively). In this case, the shoot assumes its stress-free shape. In the first scenario studied here, we also neglect the axial growth of the shoot and focus on curvature generation through tissue growth and remodeling. Thus, the shoot has a constant length (we address elongation at a later stage).% $\ell \equiv 1$, taken as a reference length unit.

\begin{figure}[ht!]
    \centering
    \includegraphics[width=.4\linewidth]{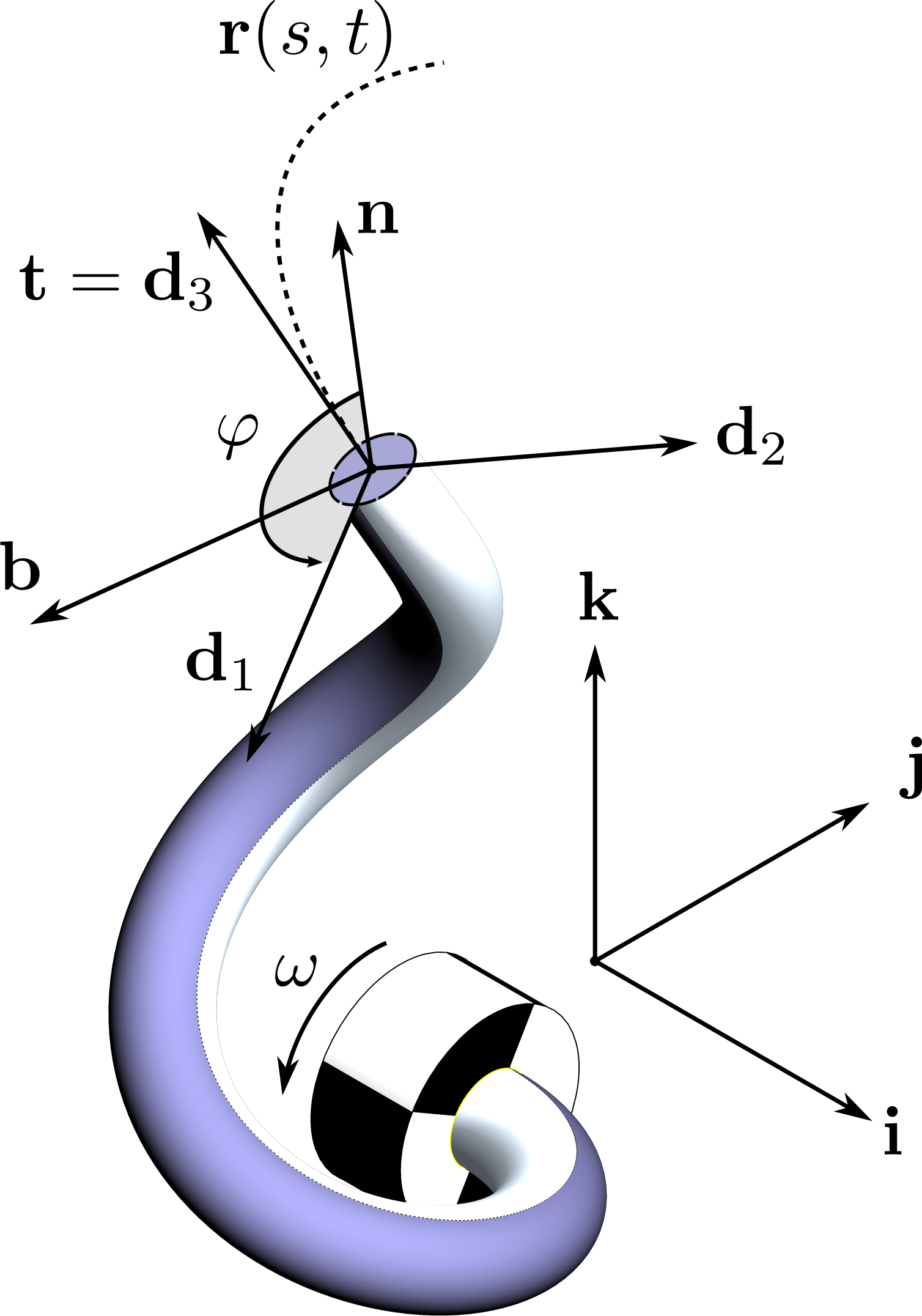}
    \caption{We model the shoot as a rod with centerline $\vec r$ and tangent $\vec t=\vec d_3$. At a given point at arclength $s$ from the base,  the vectors $\vec d_1$ and $\vec d_2$  lie in the principal directions of the cross-section. At the base of the rod, $\vec d_1$ and $\vec d_2$ are rotating around $\vec i$ with angular speed $\omega$.}
    \label{fig:fig2}
\end{figure}

\paragraph*{Model. --} The centerline of a rod is a spatial curve $\vec r\pp{s,t}=x\pp{s,t}\vec i + y\pp{s,t}\vec j + z\pp{s,t}\vec k$, parameterized here by its arclength $s\in \bb{0,\ell}$ ($s=0$ at the base) at  time $t\geq 0$; where $\left\{ \vec i,\vec j, \vec k \right\}$ is the canonical basis of $\mathbb R^3$, with $\vec k$ pointing upward against the gravity direction (\cref{fig:fig2}). The Frenet-Serret frame $\left\{ \vec{t},\vec{n},\vec{b}\right\}$, is built from the tangent vector $\vec{t} \eqdef \mathbf r '$ and the unit normal and binormal vectors,  $\vec n $ and $\vec b$,  defined through 
\begin{equation}\label{eqn:frenet}
\vec{t}' = \kappa \vec{n} , \quad 
\vec{n}' =  \tau \vec{b}  -\kappa \vec{t}, \quad
\vec{b}' = -\tau \vec{n},
\end{equation}
where $\kappa $ and $\tau $ are the curvature and torsion, respectively. In addition to its centerline, a rod is equipped with a right-handed orthonormal director basis $\vec d_1\pp{s,t}$, $\vec d_2\pp{s,t}$, and $\vec d_3\pp{s,t}=\vec t\pp{s,t}$ \citep{Goriely2017} that obeys %For an unshearable rod, we may choose $\vec d_3 = \vec{t}$. A complete kinematic description of the frame is given by
\begin{equation}\label{eqn:kinematics}
    \vec d_i'= \mathbf u \times \vec d_i, 
\quad
    \vec{\dot{ d_{\normalfont \textit i}}} = \vec w \times \vec d_i, \quad i=1,\,2,\,3.
\end{equation} 
The Darboux vector $\vec u$ and spin vector $\vec w$ obey the compatibility condition
\begin{equation}
    \vec {\dot u} - \vec w' = \vec w \times \vec u.\label{eqn:compatibility}
\end{equation}

In gravitropism, gravisensing mechanisms activate  pathways that result in differential growth of the cells \citep{KUTSCHERA2001851,blancaflor2003plant,morita2010directional,jonsson2023multiple,moulia2019posture,chauvet2019revealing,levernier2021integrative,moulia2022shaping}. Changes in curvature then occur when % then generated through differential growth where 
cells on the bottom side of the shoot extend faster than those on the upper side \citep{moulton2020multiscale,OREILLY20111239,jonsson2023multiple}. Assuming local growth laws for both gravitropism and autotropism leads, through dimensional reduction \citep{moulton2020morphoelastic,moulton2020multiscale}, to a generalization of the sine law that includes autotropism and three-dimensional effects (\cref{apdx:kinetics}):
\begin{equation}
    \vec  {\dot u} + \vec u \times \vec w = \alpha\,\vec t \times \vec k  - \beta\, \vec u \label{eqn:udot}.
\end{equation}
% with rate constants $\alpha$ and $\beta$.
% From local growth laws, one can derive the curvature evolution (with rate $\alpha$) through dimensional reduction \citep{moulton2020morphoelastic,moulton2020multiscale}. Note that solely accounting for the gravitropic influence does not typically result in equilibrium (this is evident from \cref{eqn:sachs} which, in general, does not support steady solutions). The addition of autotropism, as a restoring term with rate $\beta$, alleviates this shortcoming and allows for the existence of equilibrium configurations \citep{bastien2013}. Overall, the kinetics of the rod is expressed through \citep{moulton2020multiscale}
% \begin{equation}
%     \vec  {\dot u} + \vec u \times \vec w = \alpha\vec t \times \vec k  - \beta \vec u \label{eqn:udot},
% \end{equation}
% with $\sin \theta = \sgn\pp{\vec t\cdot \vec k} \abs{\vec t\times \vec k}$, and
Here, $\vec u\times \vec w$ accounts for the passive advection of $\vec u$ by the spin vector $\vec w$. The first term in the r.h.s accounts for gravitropism with rate constant $\alpha$. The second term models autotropism, with rate constant $\beta$, and leads to an exponential decay in time of the curvature in the absence of other effects. This equation reduces to the sine law in the planar case when $\beta=0$ and no rotation is imposed. The relative strength of gravitropism and autotropism is captured by the dimensionless \textit{bending number} $\lambda \eqdef  \alpha\ell / \beta$ \cite{bastien2013,chelakkot2017growth}. Moreover, given the constitutive hypothesis that the local growth of the cells is parallel to the axis, we have \cite{moulton2020morphoelastic}
\begin{equation}
    \vec u \cdot \vec t = 0. \label{eqn:constitutive}
\end{equation}
 The evolution of the tangent vector along the shoot is given by 
\cref{eqn:kinematics}:
\begin{equation}
\label{eqn:tprime}    \vec t'=\vec u \times\vec t .
    \end{equation}
   
%     % Indeed
    % \begin{equation}
    %     \pdiff{\vec u\cdot \vec t}{t} = \pp{\alpha\vec t \times \vec k  - \beta \vec u  + \vec w \times \vec u}\cdot \vec t + \vec u \cdot \pp{\vec w \times \vec t} = 0
    % \end{equation}
    % Note also that the multiscale morphoelastic rod model on which we base our study implies that $\vec u \cdot \vec t = 0$.
  % 
    %The system (\ref{eqn:compatibility})--(\ref{eqn:tprime}) is a
\cref{eqn:compatibility,eqn:udot,eqn:tprime,eqn:constitutive} form a closed system for $\vec u$, $\vec w$ and $\vec t$ which, given appropriate initial and boundary conditions, fully captures the shape and evolution of the shoot. For comparison, our model is the \textit{three-dimensional}, \textit{nonlinear}  generalization of the standard `AC model', which has been validated experimentally in numerous genera  \citep{bastien2013}. In particular, our approach is general enough to include complex movements such as clinostatting, enforced through a non-zero spin $\vec w\pp{0,t}\neq \vec 0$ at the base. 
    
\begin{figure}[ht!]
    \centering
    \includegraphics[width=.7\linewidth]{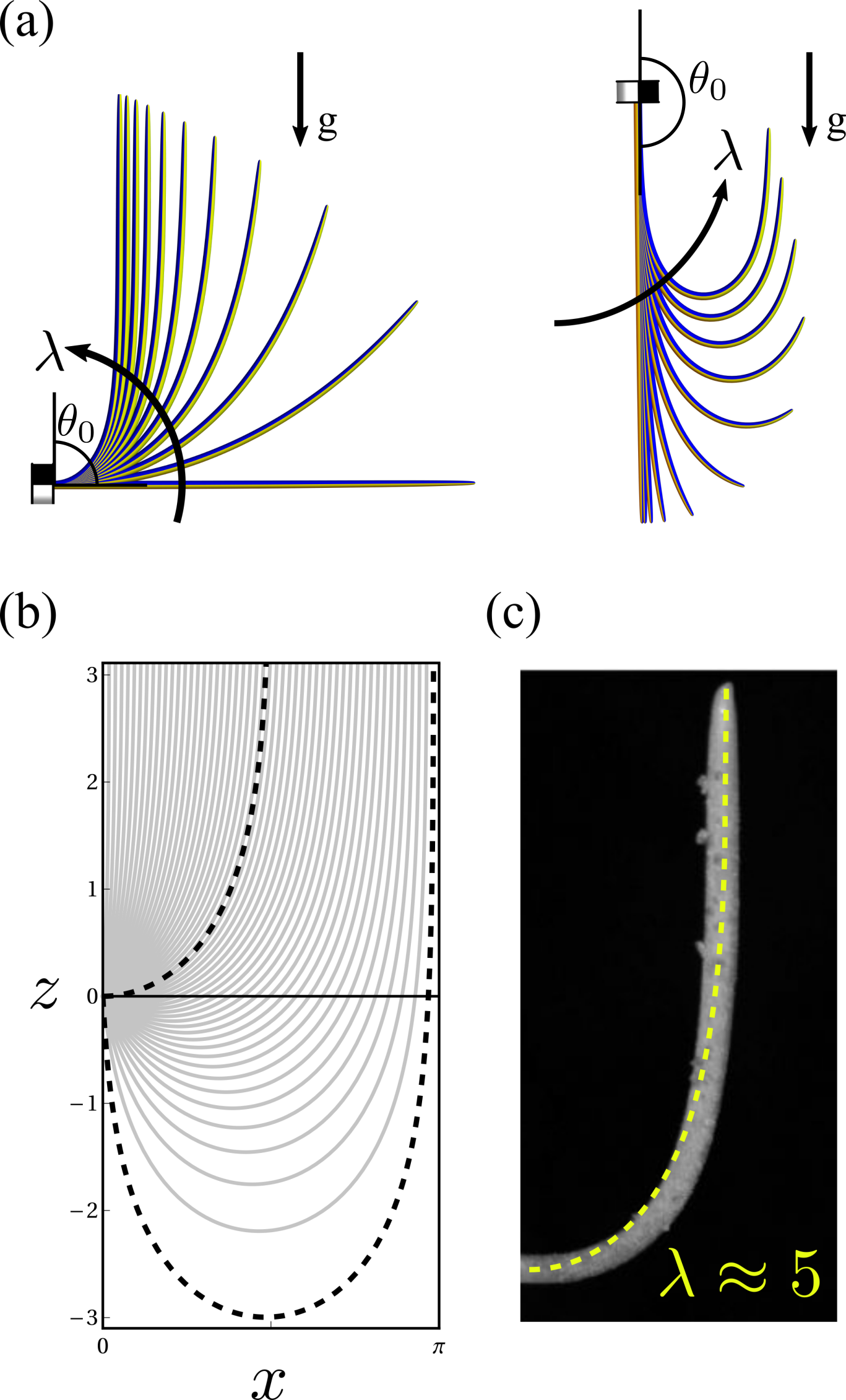}
    \caption{Steady solutions in the absence of clinostatting. (a) Horizontal clamp $\abs{\theta_0}=\pi/2$ and upside-down clamp $\abs{\theta_0}\rightarrow\pi^-$ %(with a small deviation from the vertical to break rotational symmetry) 
    for various $\lambda$. (b) The equilibrium solution is a simple caulinoid [\cref{eqn:caulinoid}] parameterized by $\theta_0$. Dashed lines show the horizontal and upside-down solutions. (c) Shape adopted by a wheat coleoptile (adapted from \cite{bastien2013}, with courtesy from B. Moulia) with overlaid caulinoid.}
    \label{fig:fig3}
\end{figure}

\paragraph*{Equilibria. --} We start our analysis by looking for equilibrium solutions in the absence of rotation, but for an arbitrary orientation of the base $\theta_0$ (\cref{no-rotation}). In that case, the equilibrium solution is planar with the exact solution $
{ \tilde z  \lambda }/{ \ell } = \log (\sin \theta_0)   - \log(\sin (\theta_0 -   { \tilde x  \lambda }/{ \ell } ) )$, 
for $0\leq  { \tilde x  \lambda }/{ \ell } < \theta_0$,
with the tilde denoting quantities at equilibrium.
% For the sake of pedagogy, we initially disregard clinostatting and focus on a non-rotating plant at equilibrium. %Therefore, setting $\vec{ \dot u}=\vec w=\vec 0 $ in \cref{eqn:udot} and using \cref{eqn:tprime} %the Darboux vector satisfies $\boldsymbol {\sf u} = - \lambda \mathbf P  \boldsymbol {\sf k} $, where $\lambda\eqdef\alpha/\beta$ is the only parameter that governs the shape.
% yields
% % \begin{equation}
% % \matr D' =- \lambda \matr D \,\mathcal L \matr D. \label{eqn:steadysol}
% % \end{equation}
% \begin{equation}
% \vec t' = \lambda\pp{\vec t \times \vec k} \times \vec t \label{eqn:steadysol}
% \end{equation}
% with $\lambda \eqdef \alpha/\beta$, the equilibrium constant of the system. 
% Here, the shoot lies in a vertical plane, and an exact solution is derived for any orientation of the base $\theta_0$ (\smref), in terms of $\lambda\eqdef \alpha/\beta $, the inverse \textit{bending number} \cite{bastien2013}. 
 We will establish that this solution is stable and gives the asymptotic shape of the shoot centerline when the base is tilted to an angle $\theta_0$ from the vertical, as shown in \cref{fig:fig3}(a).
% \todo{AG Add $\abs{\theta_0}=\pi/2$ (horizontal clamp); and $\abs{\theta_0} \rightarrow \pi^-$ (upside down clamp) on the figure. Not interesting in text. Since the file is png, I cannot edit it easily}
%The latter is shown in \cref{fig:fig3}(a) for different $\lambda$, and for two types of setup: $\abs{\theta_0}=\pi/2$ (horizontal clamp); and $\abs{\theta_0} \rightarrow \pi^-$ (upside down clamp). %In the latter case, rotational symmetry demands that the unique solution is a straight, upside-down rod. In practice, this solution will vanish in favour of a nontrivial, bent-upward solution, as soon as $\vec d _3\pp{0}$ is perturbed from the vertical; in other terms, the upside-down straight configuration is unstable with respect to perturbations of the base orientation. Therefore we consider instead a perturbed base orientation $\vec d_3\pp{0} =   - \vec k + \epsilon \vec i + \mathcal{O} \pp{\epsilon^2} $, with $0<\epsilon\ll 1$. 
On rescaling all lengths by the \textit{auto-gravitropic length} $\ell_\text{ag}\eqdef \ell/\lambda$, we obtain a universal curve [see \cref{fig:fig3}(b)]:
\begin{equation}
    \tilde z\! =\! \log (\sin \theta_0  )\! -\! \log(\sin (\theta_0\! -\!  \tilde x)), \ \quad 0\leq \tilde x < \theta_0. \label{eqn:caulinoid}
\end{equation}
We refer to this curve as the \textit{simple caulinoid} (from Latin \textit{caulis}, meaning stem).
%\cref{eqn:caulinoid} describes a universal curve, shown in \cref{fig:fig3}(b) for different $\theta_0$. 
% \begin{equation}
%    z = \log \pp{\frac{\sin \theta_0}{\sin (\theta_0 -  x)} } , \quad 0\leq x < \theta_0 <\pi. \label{eqn:caulinoid}
% \end{equation}
% with $ \quad 0\leq x < \theta_0 <\pi$  [\cref{fig:fig2}(c)]. Surprisingly, to our knowledge, this \textit{exact} solution has not been hitherto established and
% we propose the term \textit{caulinoid} as a fitting name for these universal curves (from Latin \textit{caulis}, meaning \textit{stem}). %For instance, in the horizontal setup (the most classically studied), the steady shoot shape is a portion of the caulinoid given by $
% z =\log \left(\sec x\right)$, with  $ 0\leq x < {\pi}/{2}$. %A general expression, for any orientation, is given in \smref.

\begin{figure*}[ht!]
    \centering
    \includegraphics[width=.9\linewidth]{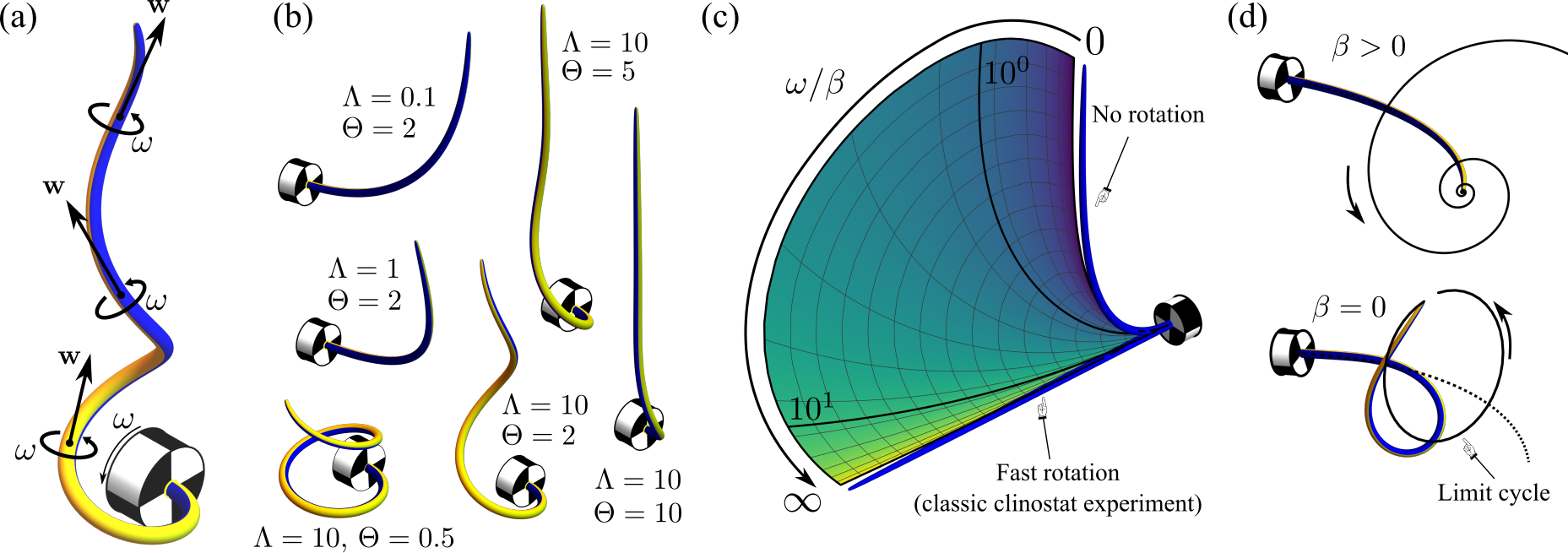}
    \caption{Dynamic equilibrium of a rotating plant. (a) The material revolves at angular speed $\omega$ around a fixed centerline. (b) Example equilibrium configurations obtained for various values of $\Lambda$ and $\Theta$. (c) Dependency of the equilibrium solution on $\omega$. Blue rods show the limits $\omega\rightarrow 0$ (no rotation) and $\omega \rightarrow \infty$ (standard clinostat experiment). The surface shows the set of equilibrium solutions obtained for finite values of $\omega/\beta$ ($\lambda=5$). (d) Course of the apex in two cases, $\beta>0$, with  convergence to equilibrium ($\beta=\omega/5$), and $\beta = 0 $, after convergence to a limit cycle (in both cases $\alpha=\omega$). Dashed line shows the corresponding equilibrium solution.}
    \label{fig:fig4} 
\end{figure*}

% \paragraph{Rotating clinostat. --} 
Next, we consider a clinostat imparting a counterclockwise rotation around the horizontal axis $\vec i$ with period $T=2\pi/\omega$. In this case, the boundary conditions are $\vec t\pp{0,t} = \vec i$ and $\vec w \pp{0,t} =\omega \vec i$. By definition, at equilibrium, we have $\vec{\dot w} = \vec {\dot u} = \vec{\dot t}=\vec 0$, which gives $\vec w = \omega \vec t$. In this configuration, the shoot revolves at constant angular velocity $\omega$ about a \textit{fixed} centerline [\cref{fig:fig3}(a)] with tangent vector given by (\cref{sec:equilibrium-clinostat})
\begin{align}
  \vec{\tilde t}\pp{s} =& \frac{\cos \Lambda s }{\cosh \Theta s}  \,  \vec i -\frac{\sin \Lambda s }{\cosh \Theta s} \,\vec j  +  \tanh (\Theta s) \,\vec k,\label{eqn:static-tau}
\end{align}
where $\Lambda\eqdef {\alpha\omega}/\pp{\omega^2+\beta^2}$ and $\Theta\eqdef{\alpha\beta}/\pp{\omega^2+\beta^2}$. The curvature, $ \tilde\kappa\pp{s} = \sqrt{\Theta ^2+\Lambda ^2} \,  \sech\Theta  s $, and  torsion, $\tilde\tau\pp{s} = -\Lambda \tanh \Theta s$, of this general caulinoid satisfy% there is a first integral %$\mathcal E\pp{\kappa, \tau}$ such that
\begin{equation}
 \frac{\tilde\kappa^2}{\Theta ^2+\Lambda ^2}+\frac{\tilde\tau^2}{\Lambda ^2}=1.\label{eqn:firstintegral}
\end{equation}
Thus, along an equilibrium solution, starting from $\tilde\tau\pp{0} = 0$ at the base, the torsion increases while the curvature decreases along an ellipse in the curvature-torsion plane. In physical space, the centerline follows a modulated left-handed helix that gradually uncoils away from the base towards the vertical, and we can interpret $\Lambda$ and $\Theta$ as the curve's \textit{winding} and \textit{rise} densities [\cref{fig:fig4}(b)]. In the limit $\omega\rightarrow 0$, we have $\Lambda = 0$ and $\Theta=1/\ell_\text{ag}$, recovering the planar case discussed above. When $\omega \rightarrow \infty $, the plant remains straight with  $\Lambda=\Theta=0$  [\cref{fig:fig4}(c)]. The equilibrium curve is uniquely determined by $\Lambda$ and $\Theta$. Experimentally, given  $\omega$, both parameters $\alpha$ and $\beta$ can thus be estimated uniquely from the centerline (unlike in the planar case), e.g. by using the height of the plant $ H=\tilde{z}(1)=\log\pp{\cosh \Theta}/\Theta$ and the radius of the caulinoid at the base $R=1/\Lambda$.

A numerical linear stability analysis of the full system (\cref{apdx:stability}) conducted across a wide range of realistic parameters $\lambda \in \bb{0.1, 100}$ [\cref{fig:fig3}(c)], consistent with reported values \citep{bastien2013,chelakkot2017growth,tsugawa2023shoot}, reveals that, for $\beta>0$, the equilibrium solution is linearly stable. Further, the local dynamics near the base can be obtained asymptotically, showing that the Darboux vector spirals towards its equilibrium value with a typical exponential decay $\E^{-\beta t}$ [see \cref{fig:fig4}(d) and Movie 1].
%, the norm of $\vec u(0,t)-\vec{\tilde{u}}(0)$ decaying as   (\smref{} and Movie 1). This observation suggests that the equilibrium is a global attractor. This rate of convergence was further supported by numerical linear stability analysis of the full system, conducted across a wide range of realistic parameters $\lambda \in \bb{0.1, 100}$ [\cref{fig:fig3}(c
%)] consistent with reported values \citep{bastien2013,chelakkot2017growth}. 
In the limit case $\beta= 0$ but with $\omega\not=0$ \citep{moulton2020multiscale}, the equilibrium solution is a segment of a horizontal circle of radius $\omega/\alpha $. Here, however, the previous stability result does not apply and the shoot orbits around the equilibrium [see \cref{fig:fig4}(d) and Movie 2].

\begin{figure}[ht!]
    \centering
    \includegraphics[width=\linewidth]{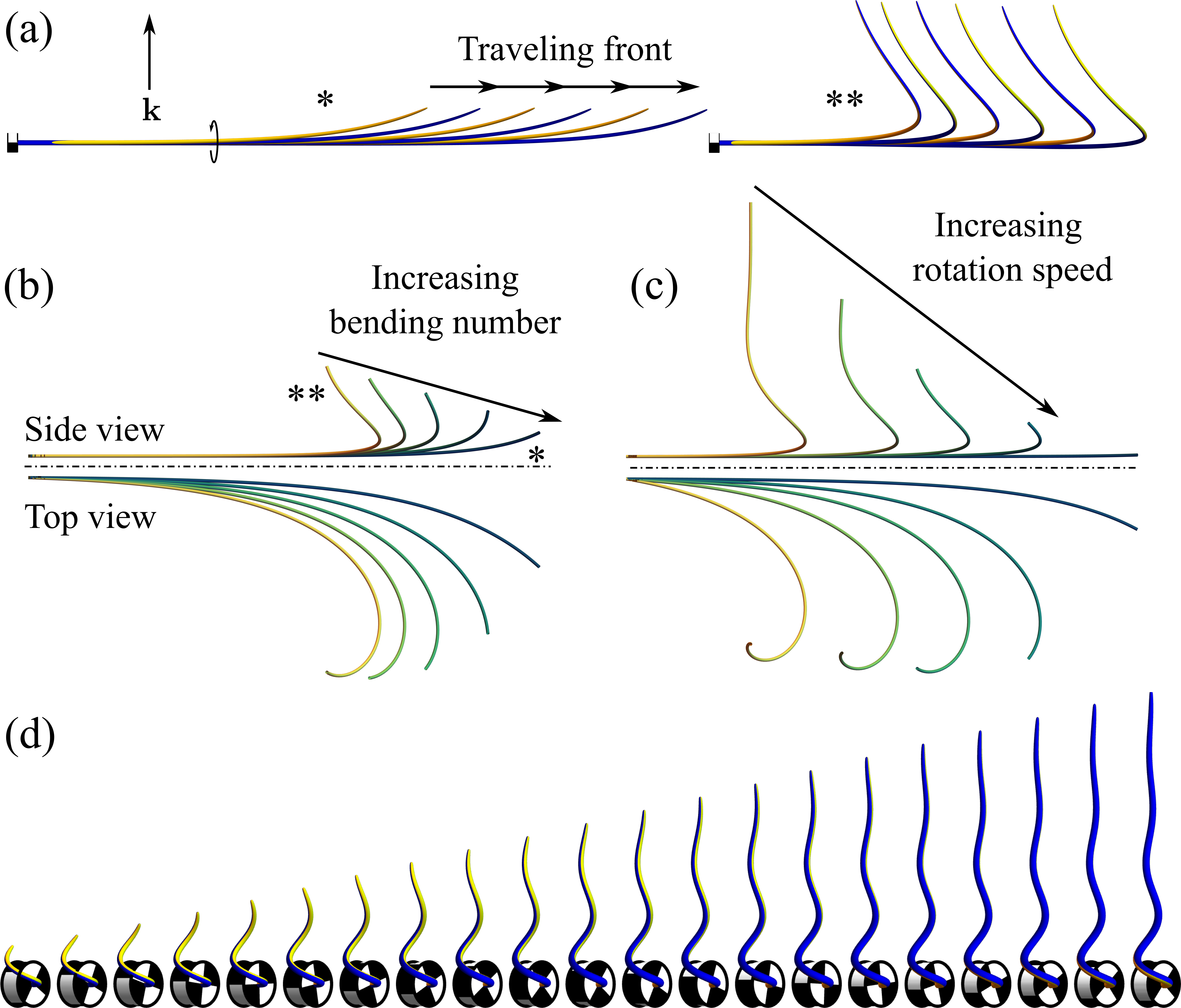}
    \caption{Growth. (a) Examples of simulated growing shoots (parameters: $\alpha=\omega$ and $5 \omega$; $\beta=\omega$; $\delta = \ell$; $\Gamma_0=\omega/10$). (b, c) Solitary wave profiles computed for  different (b) bending numbers $\lambda$; and (c) rotation speeds $\omega$. The labels $*$ and $**$ indicate corresponding sets of parameters between the simulation and the asymptotic profile. (d) Uniform growth rate ($\delta\gg \ell$): The plant spreads along a unique caulinoid (here, $\Lambda=5$, $\Theta=1$, $\Gamma_0=\omega/10$).}
    \label{fig:fig5}
\end{figure}

\paragraph*{Shoot elongation. --}
Plants also lengthen due to the coordinated expansion of the cells along the central axis. Generally, this primary growth is mostly confined to a region close to the apex \citep{SILK1979481}. To model elongation, including %clinostatic rotation and 
apical dominance, we assume that both the  tropic response and axial growth gradually diminish as we move away from the apex with exponential decay of  characteristic length $\delta$ and with growth $\Gamma_0$ and auto-gravitropic rates, $\beta$ and $\alpha$, at the tip (\cref{growth}). In this case, the system supports a traveling front solution connecting a flat base to a steady apical structure migrating forward at a speed $c=\Gamma_0\delta$ [see \cref{fig:fig5}(a) and Movie 3]. The shape of this solitary wave can be described in terms of an initial value problem that can be integrated numerically. \cref{fig:fig5}(b) shows example solutions obtained for various rotation speeds and bending numbers $\lambda$. An interesting limit is $\ell \ll \delta $ (uniform growth rates along the shoot). Assuming a timescale separation  $\beta \gg \Gamma_0$, and noting that $\Lambda$ and $\Theta$ are independent of $\ell$, we see that the shoot's shape will progress quasi-statically, spreading itself uniformly along a unique caulinoid [see \cref{fig:fig5}(d) and Movie 4]. The existence of these solutions demonstrates that steady configurations are a robust property of the system that can persist even upon significant elongation.

\paragraph*{Discussion. --}
The clinostat holds a significant place in plant physics, addressing a precise technical challenge: simulating weightlessness by effectively `confusing' the plant through fast rotation. At lower speeds, the interaction between rotation,  gravitropism, and autotropism reveals more subtle behaviors. 
%By extending the paradigm model of \textit{gravi-proprioception} \citep{bastien2013} to three-dimensional rotations, we have investigated the gravitropic response under all rotation regimes. 
A distinct property of this system is the universal existence of a \textit{dynamic} equilibrium where the shoot revolves around a steady centerline, the caulinoid. %\cref{eqn:static-tau}. 
This equilibrium is dynamic as it requires cyclic deformations in the material to maintain this configuration as rotation is applied. In contrast to the classic planar case, whose equilibrium is determined solely by $\lambda $ (\cref{fig:fig3}), this solution is \textit{uniquely} characterized through two dimensionless numbers $\alpha\ell/\omega$ and $\beta/\omega$. %In other words, the equilibrium shape unequivocally incorporates all the physiological parameters of the system.
When the plant undergoes elongation, two distinct behaviors emerge: solitary waves when growth, autotropism and gravitropism are confined to the tip; or stationary elongation along a unique caulinoid when the shoot grows uniformly. In conclusion, we predict that a clinostatting shoot will naturally assume the sole shape that enables it to counterbalance rotation and minimize its overall movement in the laboratory frame, strikingly, even in the absence of a dedicated rotation-sensing mechanism. 
%\todo{Variational principle?}

 The importance of proprioception in plant posture control is now well established  \citep{bastien2013,hamant_how_2016,moulia2019posture,riviere2020hook,moulia2021fluctuations,moulia2021fluctuations,moulia2022shaping}. We further showed that the role of proprioception, in the form of autotropism, is crucial in stabilizing the clinostatting shoot, as its absence would lead to non-steady behaviors \citep{moulton2020multiscale}. Physically, autotropism acts as a damping mechanism in curvature space, hence providing a stabilization mechanism.
 % Notably, \citeauthor{bastien2013} \citep{bastien2013} have systematically evidenced the vital importance of proprioception in gravitropic straightening. Here, we demonstrate that convergence towards equilibrium is also predicated upon autotropism; the absence thereof leads to non-steady behaviors. %Our model can be readily extended to phototropism, mechanical effects such as self-weight, richer patterns of rotation, or a combination of all. 
The exact caulinoid solutions may be difficult to observe experimentally with precision as it would require pristine conditions. Further, in plants, heterogeneity, stochasticity and other tropic responses also play a role. Yet, these ideal solutions  present a new paradigm for the study of plant shapes and the design of experiments. They can be further generalized to include other effects, such as light, or elasticity \citep{moulton2020multiscale}. They demonstrate that the coupling of internal and external stimuli is key for shape control, a problem of general importance in biology with direct implications for non-living active materials.
% Overall, our study provides a paradigmatic framework for deciphering the mechanisms through which a plant develops robustly under changing environmental conditions. This is a problem of general importance in morphogenesis, agronomy, and ecology, with direct implications for non-living active materials. %, and opens an avenue of experimental and theoretical research. 
\newline

A.G. acknowledges support from the Engineering and Physical Sciences Research Council of Great Britain under Research Grant No. EP/R020205/1. H.A.H. acknowledges support from the Royal Society under University Research Fellowship No. {URF/R/211032}. For the purpose of Open Access, the author has applied a CC BY public copyright license to any Author Accepted Manuscript (AAM) version arising from this submission.

\appendix

\section{Kinetics of curvature evolution\label{apdx:kinetics}}

The auto-gravitropic governing law, derived in \citep{moulton2020multiscale}, reads in vector form:
 \begin{equation}
    \boldsymbol {\dot{\sf u}} = \alpha \boldsymbol{\sf t} \times \boldsymbol{\sf k} - \beta\boldsymbol{\sf u}.\label{eqn:localudot}
\end{equation}
Here, we have used Antman's {\textsf{sans-serif}} notations \citep{antman2005problems} to denote vector field attached to a curve and expressed in the local material frame $\vec d_i$, i.e. for a vector field  $\vec u \pp{s,t}$, we write
\begin{equation}
\vec u \pp{s,t}= \sum _{i=1}^3 {\sf u}_i \pp{s,t} \vec d_i \pp{s,t}, \label{eqn:sfu}
\end{equation} and $\boldsymbol {\sf u} = \pp{{\sf u}_1,{\sf u}_2,{\sf u}_3}$ then denotes the vector of local coordinates ${\sf u}_i = \vec u \cdot \vec d_i $ (in particular, $\vec t=\vec d_3$ implies $\boldsymbol{\sf t}=\pp{0,0,1}$). Thus, \cref{eqn:localudot} expresses the evolution of curvatures from a local point of view, i.e. in a reference frame attached to the material. In our case, since gravity is important, it is convenient to express the dynamics in the non-rotating, laboratory frame (indeed, the equilibrium solutions are naturally expressed in the laboratory frame). Therefore, we differentiate \cref{eqn:sfu} with respect to time, and use $\vec{\dot d }_i = \vec w\times \vec {d}_i$ [\cref{eqn:kinematics}], to obtain the kinematic relation 
\begin{equation}
    \vec{\dot u} + \vec u \times \vec w = \sum_{i=1}^3 \dot{\sf u}_i \vec d_i. \label{eqn:changecoordinate}
\end{equation}
Using the rotational invariance of the cross product, \cref{eqn:localudot,eqn:changecoordinate} directly provide the expression given in \cref{eqn:udot} for $\vec {\dot u}$.

\section{Equilibrium solutions\label{apdx:equilibria}}

\subsection{Without rotation\label{no-rotation}}
We derive the equilibrium solutions for the non-rotating case ($\omega=0$). Here, we choose $\ell\equiv 1$ as a reference length unit. Setting $\vec{\dot u}=\vec 0$ in \cref{eqn:udot} provides $
    \vec {\tilde u} = \lambda \vec {\tilde t}\times \vec k$, which can be substituted into \cref{eqn:tprime} to obtain 
    \begin{align}        
    \vec {\tilde t}' = \lambda \pp{\vec {\tilde t}\times \vec k}\times \vec {\tilde t}.   \end{align}
    Provided an initial tilt $0\leq \theta_0 <\pi$, such that $\vec {\tilde t}\pp{0} = \sin\theta_0 \,\vec i + \cos \theta_0 \,\vec k$, we  integrate this equation and derive the tangent
\begin{align}
    \vec{\tilde  t}\pp{s} &=\frac{\sin \theta_0}{\cos \theta_0 \sinh \lambda s+\cosh \lambda s} \,\vec i\nonumber\\&+{\frac{(\cos \theta_0+1) \E^{2 \lambda  s}-1+\cos\theta_0}{(\cos \theta_0+1) \E^{2 \lambda  s}+1-\cos\theta_0}}\,\vec k.
\end{align}
Integrating once more gives the position vector $\vec{\tilde  r }\pp{s} = \tilde x\pp{s}\,\vec i + \tilde z\pp{s}\,\vec k $:
\begin{subequations}
\begin{equation}    \lambda \tilde x\pp{s} =  {\theta_0 -2 \arccot\left( \E^{\lambda  s}\cot \frac{\theta_0 }{2}\right)},\label{eqn:xs}
\end{equation}
\begin{equation}
     \lambda  \tilde z\pp{s}= \log \bb{1 +\cos^2\pp{{\theta_0}/{2}} \pp{\E^{2 \lambda  s}-1}}- \lambda s.
\end{equation}
\end{subequations}
Inverting \cref{eqn:xs} and rescaling all lengths as $\tilde x\rightarrow\tilde  x/ \lambda$, $\tilde z\rightarrow \tilde z/ \lambda$, we obtain an implicit relation between $\tilde z$ and $\tilde x$ [\cref{eqn:caulinoid}], which corresponds to a universal equilibrium shape for all orientations $\theta_0$ of the shoot.

\subsection{With rotation\label{sec:equilibrium-clinostat}}
\begin{figure}[ht!]
    \centering
    \includegraphics[width=.95\linewidth]{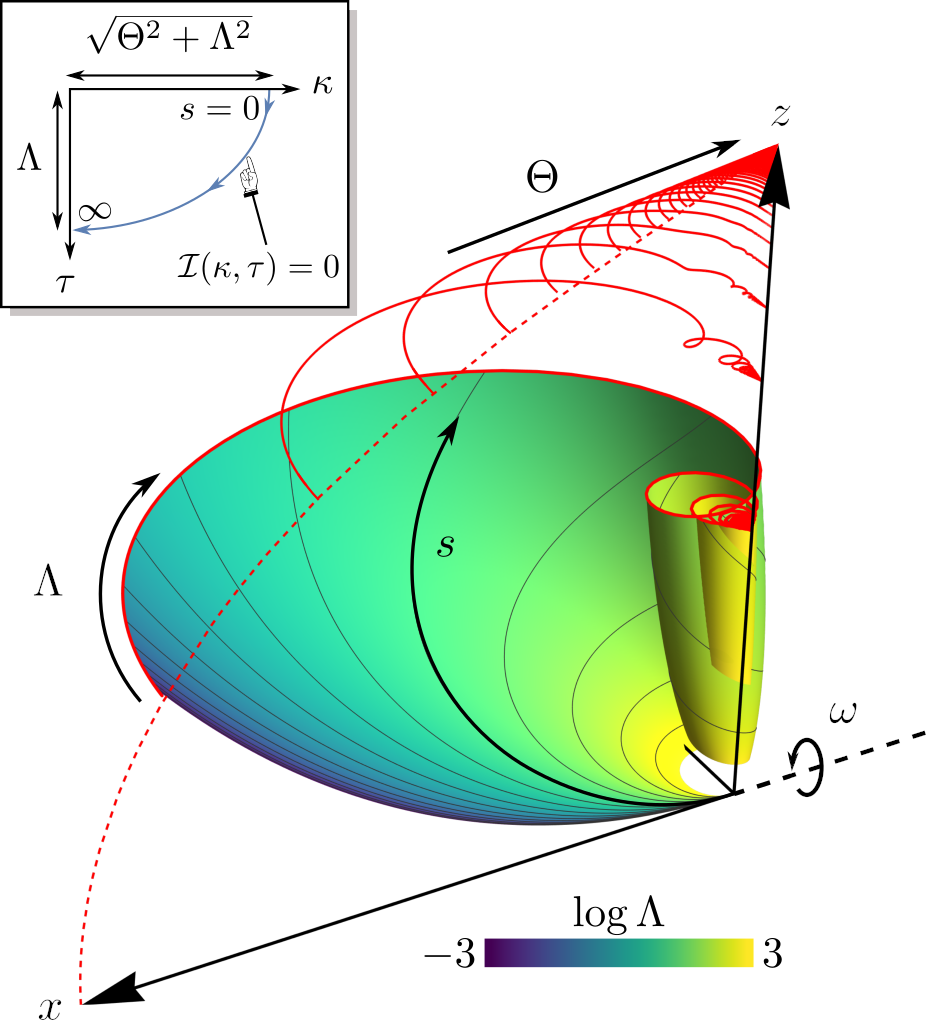}
    \caption{Set of equilibrium solutions. The colored surface plot sweeps solutions for a range of $\Lambda$ (with $\Theta=1$). Red solid lines show the course of the shoot tip $s=0$ as $\Lambda$ varies, and for different values of $\Theta>1$, with height given by $h\pp{\Theta}=\log (\cosh \Theta )/\Theta$. Red dashed line shows the tip position for $\Lambda =0$ as a function of $\Theta$, given by $\pp{x\pp{\Theta}, z\pp{\Theta}}=\pp{\gd\Theta/\Theta ,h\pp{\Theta}}$. Inset shows the path of the solution in the $\kappa$--$\tau$ plane [\cref{eqn:firstintegral}].}
    \label{fig:nonstaticsol}
\end{figure}

Next, we derive the equilibrium solution for a plant undergoing rotation ($\omega>0$). To determine the equilibrium shape, we posit $\vec {\dot w} = \vec{\dot t} = \vec {\dot u} = \vec 0$. \cref{eqn:kinematics,eqn:compatibility} directly provide that $\vec {\tilde w}= \omega\vec {\tilde t}  $. Substituting this ansatz into \cref{eqn:udot}, we obtain 
\begin{align}
\beta  \vec {\tilde u} = \vec {\tilde t}\times\pp{\alpha\vec k + \omega \vec {\tilde u}}.    
\end{align} 
On inverting this identity, we can express $\vec {\tilde u}$ as an explicit function of $\vec {\tilde t}$, given by
\begin{align}\nonumber
   \vec{\tilde u } = \pp{ \Lambda  \tilde t_1 \tilde t_3+\Theta   \tilde t_2}\,\vec i &+ \pp{\Lambda  \tilde t_2 \tilde t_3-\Theta  \tilde t_1}\,\vec j\\ &+\Lambda  \pp{\tilde t_3^2-1} \,\vec k, \end{align}
with $\vec {\tilde t} = \tilde t_1 \,\vec i + \tilde t_2\,\vec j + \tilde t_3\,\vec k$.  Substituting this last expression into \cref{eqn:tprime} and integrating it, we obtain the expression for the tangent given by \cref{eqn:static-tau}. Remarkably, we can integrate the tangent to obtain an exact parameterization of the centerline $\vec{\tilde r}$,  in terms of the hypergeometric function $_2F_1$, the harmonic number $H_n$ and the polygamma function of order zero $\psi^{(0)}$:
\begin{widetext}
\begin{subequations}
        \begin{align}
         \tilde   x\pp{s} &= \frac{\E^{s (\Theta +\I \Lambda )}}{\Theta +\I \Lambda } \, _2F_1\left(1,\frac{\Theta +\I \Lambda }{2 \Theta };\frac{3\Theta+\I \Lambda }{2 \Theta };-\E^{2 s \Theta }\right) +\frac{\E^{s (\Theta -\I \Lambda )} \, }{\Theta -\I \Lambda } \, _2F_1\left(1,\frac{\Theta -\I \Lambda }{2 \Theta };\frac{3\Theta -\I \Lambda }{2 \Theta };-\E^{2 s \Theta }\right)\nonumber\\&-\frac{\pi }{2 \Theta } \sech\left(\frac{\pi  \Lambda }{2 \Theta }\right),\\
           \tilde y\pp{s} &= \frac{\I }{4 \Theta }  \bb{\psi ^{(0)}\left(\frac{\Theta +\I \Lambda }{4 \Theta }\right)-\psi ^{(0)}\left(\frac{3\Theta+\I \Lambda }{4 \Theta }\right)+H_{-\frac{\Theta +\I \Lambda }{4 \Theta }}-H_{-\frac{3\Theta+\I \Lambda }{4 \Theta }}  }\nonumber\\&+\frac{\E^{s (\Theta +\I \Lambda )}  }{\Lambda -\I \Theta }\,_2F_1\left(1,\frac{\Theta +\I \Lambda }{2 \Theta };\frac{3\Theta+\I \Lambda }{2 \Theta };-\E^{2 s \Theta }\right)+\frac{\E^{s (\Theta -\I \Lambda )}}{\Lambda +\I \Theta } \, _2F_1\left(1,\frac{\Theta -\I \Lambda }{2 \Theta };\frac{3\Theta-\I \Lambda }{2 \Theta };-\E^{2 s \Theta }\right),\\
           \tilde z\pp{s}& = \frac{\log (\cosh (\Theta  s))}{\Theta }.
        \end{align}
\end{subequations}
\end{widetext}
\cref{fig:nonstaticsol} shows the set of solution shapes for different values of $\Lambda$ and $\Theta$. Inset shows the path of the solution in the $\kappa$-$\tau$ space, which follows the ellipse given by \cref{eqn:firstintegral}.

\section{Numerical resolution of the nonlinear system\label{nonlinear-numerics}}

We use a  method based on Chebyshev polynomials to integrate numerically the nonlinear system given by \cref{eqn:udot,eqn:tprime,eqn:compatibility,eqn:constitutive}. We first remark that the system, albeit originally defined for $s\in \bb{0,1}$, can be extended naturally to $s\in\bb{-1,1}$ (by considering two `twin' shoots oriented opposite to each other with respect to the plane $y$-$z$). Here, the extended equilibrium solution is invariant with respect to the mirror symmetry $x\rightarrow - x$, $s\rightarrow-s$. This situation is ideal for using  Chebyshev polynomials of the first kind $T_n$ \cite{Hale2015} as they are defined canonically on $\bb{-1,1}$. Thus, we consider the truncated Chebyshev expansions for the variables
\begin{subequations}
\label{eqn:chebyshev-expansion}
    \begin{equation}
        \vec t \approx \sum_{n=0}^N \vec T^n T_n,
    \end{equation}
        \begin{equation}
        \vec w \approx \sum_{n=0}^N \vec W^n T_n,
    \end{equation}
        \begin{equation}
        \vec u \approx \sum_{n=0}^N \vec U^n T_n,
    \end{equation}
\end{subequations}
with $N$ a positive integer. 

The formal solutions for $\vec t$ and $\vec w$,
\begin{subequations}
\begin{equation}
    \vec t = \vec i + \int_0 ^s \vec u\times \vec t,
    \label{eqn:formalsol_t}
\end{equation}
\begin{equation}
    \vec w = \omega \vec i +\alpha  \int_0 ^s \vec t \times \vec k - \beta \int_0^s \vec u,
    \label{eqn:formalsol_w}
\end{equation}
\end{subequations}
can be decomposed on the Chebyshev basis as follows. From the products $T_nT_m= \pp{T_{n+m}+ T_{\abs{n-m}}}/2$ \cite{Hale2015}, 
we derive the expansion of the cross products, i.e., for any vector field $\vec a$ and $\vec b$ with respective Chebyshev coefficients $\vec A^n$ and $\vec B^n$, we have
\begin{widetext}
\begin{equation}
    \vec a\times \vec b
    =  \frac12 \sum_{p=0}^\infty \pp{\vec A^p\times \vec B^p + \vec A^p\times \vec B^{-p}}T_0 + \frac12 \sum_{n=1}^\infty\sum_{p=0}^n  \pp{ \vec A^{p}\times \vec B^{n-p}+ \vec A^{p}\times \vec B^{n+p}+\vec A^{n+p}\times \vec B^{p}}T_n .
\end{equation}
    \end{widetext}  
For integration, we use the recurrence formulae \citep{Hale2015}
\begin{subequations}
\begin{equation}
    \int T_0 =  T_1;
\quad \int T_1 = \frac14 \pp{T_2+T_0};
\end{equation}
\begin{equation}
\int T_n = \frac12 \pp{\frac{T_{n+1}}{n+1} - \frac{T_{n-1}}{n-1}},\quad \forall n\geq 2,
\end{equation}  
\end{subequations}
to obtain
    \begin{equation}
            \int \vec a  = \frac{\vec A^1}{4}T_0 + \vec A^0 T_1 +  \sum_{n=2}^\infty \frac{\vec A^{n-1}-\vec A^{n+1}}{2n}T_{n}  . \label{eqn:chebyshev-integral}
        \end{equation}
Conveniently, integration corresponds to a linear operation on the $\vec A^n$, whose matrix can be precomputed. 

Given the coefficients $\vec U^n$, the Chebyshev expansion of \cref{eqn:formalsol_t} yields a linear system that can be inverted to obtain the $\vec T^n$. Then, the $\vec W^n$ are obtained by direct integration, using \cref{eqn:chebyshev-integral}. After expressing the $\vec T^n$ and $\vec W^n$ as functions of the $\vec U^n$, we obtain a dynamical system  of the form
\begin{equation}
      {\dot {\mathbb U}}
=\mathcal F \pp{ {\mathbb U}},\label{eqn:ucoefs_dot}\end{equation}
where ${\mathbb U}$ is the $3\pp{N+1}$-dimensional vector formed by the concatenation of the $\vec U^n$; and $\mathcal F$ is a second-degree polynomial vector that is evaluated numerically. Provided appropriate initial conditions, \cref{eqn:ucoefs_dot} can be integrated numerically using a standard IVP solver (here we used \textit{Mathematica}'s built-in routine \textit{NDSolve}). 

A general problem is to find an initial condition for $\vec u$ that satisfies the orthogonality condition,  \cref{eqn:constitutive}. Indeed, by differentiating $\vec u\cdot \vec t$ with respect to time and using \cref{eqn:udot}, we observe that 
\begin{equation}
    \pdiff{}{t}\pp{\vec u\cdot \vec  t} =  - \beta \vec u\cdot \vec t. \label{eqn:udott}
\end{equation}
Since $\beta>0$, \cref{eqn:constitutive} is a stable property, in particular, if \cref{eqn:constitutive} is satisfied at $t=0$, it will be automatically satisfied at all times $t$. Note that,  if $\vec u\cdot \vec t = 0$, then we have automatically%, from $\vec t \times \vec t' = \vec t \times \pp{\vec u \times \vec t}   =  \vec u - \pp{\vec u \cdot \vec t} \vec t$,
% we have
\begin{equation}
    \vec u = \vec t \times \vec t' \label{eqn:uttprime}
\end{equation}
(the converse is trivial). Thus a suitable initial condition can always be found by first defining a curve and its tangent $\vec t$; and then obtaining $\vec u$ through \cref{eqn:uttprime}. Once an initial configuration is defined, the initial Chebyshev coefficients for $ \mathbb U\pp{0}$ are computed efficiently by means of the discrete cosine transform \citep{press2007numerical}.

\section{Stability\label{apdx:stability}}

\subsection{Asymptotic analysis near the base \label{base}}
To gain insight into the dynamics of the shoot and its stability, it is useful to first restrict our attention to the base of the plant, $s=0$, where $\vec t\pp{0,t}=\vec i$ and $\vec w\pp{0,t}=\omega \vec i$. Letting $\vec U \pp{t} = \vec u \pp{0,t}$,
\cref{eqn:udot} reduces to% a system of ordinary differential equations [\cref{eqn:udot}]
 \begin{equation}
 \dot U_2 = - \alpha - \beta U_2 - \omega U_3,\quad 
 \dot U_3  =\omega U_2 -\beta U_3 .
 \end{equation}
with $\vec U = U_2 \,\vec j + U_3 \,\vec k $. We have $U_1=0$ by \cref{eqn:constitutive}. The system admits a unique fixed point $\pp{U_2,U_3}=\pp{-\Theta,-\Lambda}$ (this is simply the equilibrium curvatures at the origin derived in \cref{sec:equilibrium-clinostat}), associated with a pair of conjugate eigenvalues $-\beta \pm \omega\I$ with negative real part: The fixed point is a spiral sink associated with a decaying amplitude $\sim \E^{-\beta t}$ and rotation speed $\omega$. When $\beta=0$ the fixed point is a center and the solution orbits around the fixed point. 

\begin{figure}
    \centering
    \includegraphics[width=\linewidth]{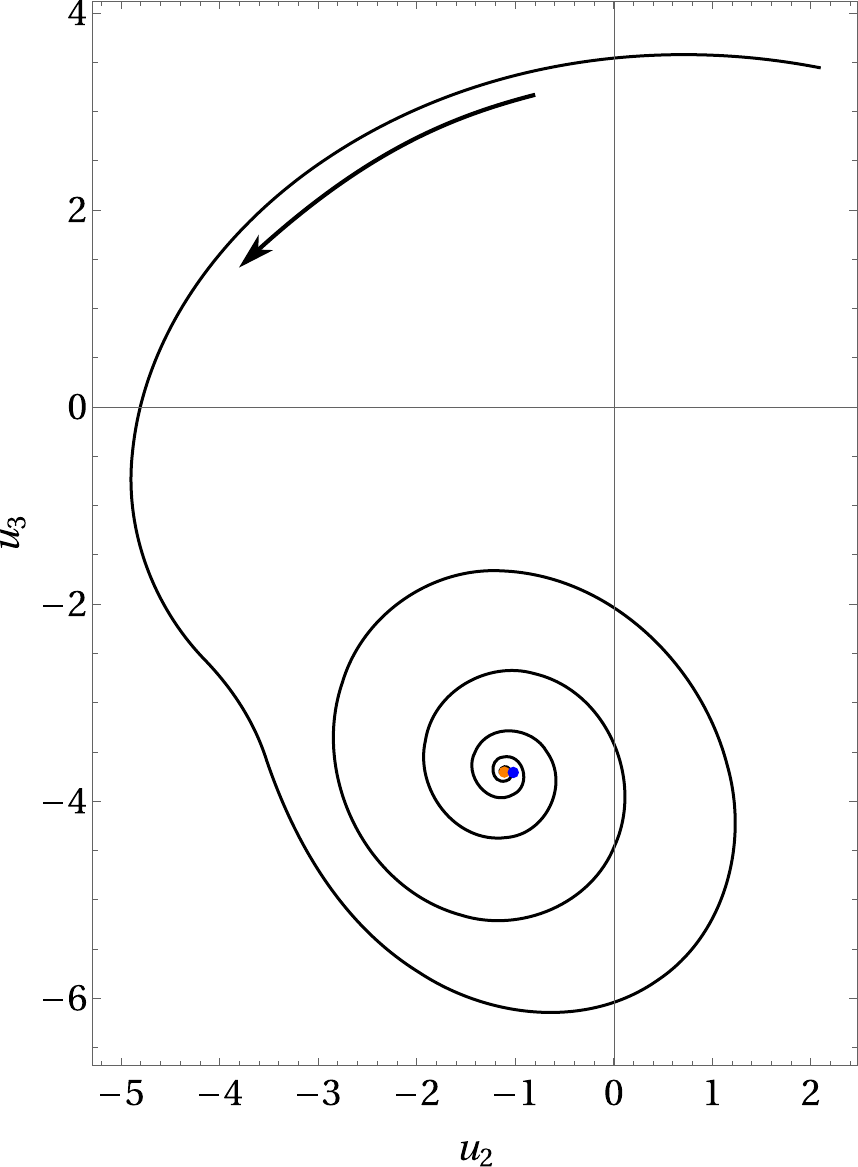}
    \caption{Example course of the Darboux vector $\vec u\pp{t}$ in the plane $\vec j$-$\vec k$, computed asymptotically (to second order in $s$) near the base ($s=0.25$). The asymptotic solution spirals towards an equilibrium value ($\alpha=4 \omega $, $\beta=0.2\omega$). Blue and orange dots show the exact value of $\vec u$ at equilibrium at $s$, and its second-order approximation, respectively.}
    \label{fig:base-solution}
\end{figure}

We can extend this analysis to higher orders in $s>0$ in principle (that is, expanding all variables in orders of $s$ and performing a regular perturbation). For instance, \cref{fig:base-solution} shows the second-order approximation of the solution taken at $s=0.25$. The second-order estimate converges towards equilibrium when $\beta>0$ and $s\ll1$ (in the case $\beta=0$ however, there is a secular term that must be treated by a dedicated method, but we leave this problem outside the scope of this study, focusing on the physiologically relevant case $\beta>0$).

\subsection{Linear stability analysis}
The previous analysis provides insight into the dynamics of the system; however, in principle, it is valid only near the base. To complement that approach, we perform a linear stability analysis of the equilibrium solution. Therefore, we take the first variation of \cref{eqn:udot,eqn:compatibility,eqn:tprime,eqn:constitutive} around the base equilibrium solution derived in \cref{sec:equilibrium-clinostat}. Rearranging the terms, we obtain:
\begin{subequations}
\label{eqn:first-variation}
\begin{equation}
    \delta \vec t' = \delta \vec u \times \vec{\tilde t} + \vec{\tilde u}\times \delta \vec t, \label{eqn:deltatprime}
\end{equation}
\begin{equation}
    \delta \vec w'= \alpha\, \delta \vec t \times \vec k - \beta \delta \vec u, \label{eqn:deltawprime}
\end{equation}
\begin{equation}
     \delta \vec{\dot u} =\delta \vec w' + \delta \vec w \times \vec{\tilde u} + \vec {\tilde w} \times \delta \vec u, \label{eqn:deltadotu2}
\end{equation}
\label{eqn:variation}
\end{subequations}
with the conditions
\begin{equation}
   \vec {\tilde u} \cdot \delta \vec t = - \delta \vec u\cdot \vec {\tilde t} ,\quad  \vec {\tilde t}\cdot \delta \vec t = 0. 
   \label{eqn:orthogonality}
\end{equation}
The boundary conditions at $s=0$ fix the values of $\vec t\pp{0,t}$ and $\vec w \pp{0,t}$, thus,
\begin{equation}
    \delta \vec t\pp{0,t} = \vec 0, \quad \delta \vec w\pp{0,t} = \vec 0. \label{eqn:deltabc}
\end{equation}

We start by solving \cref{eqn:deltatprime}. As can be seen, a linearly independent basis of solutions for the homogeneous part of \cref{eqn:deltatprime} 
is provided by the $\vec d_i $ at equilibrium (defined up to an arbitrary rotation of the clinostat). A particular solution is then obtained by means of variation of constants. For a given $\delta\vec u$, the solutions to \cref{eqn:deltatprime,eqn:deltawprime,eqn:deltabc} are:
\begin{subequations}\label{eqn:deltatwsol}
        \begin{equation}
        \delta \vec t = \vec { d}_1 \int_0^s  \delta{\vec u}\cdot \vec {  d}_2   -\vec {  d}_2 \int_0^s \delta{\vec u}\cdot \vec { d}_1,\label{eqn:deltatsol}
    \end{equation}
    \begin{equation}
        \delta \vec w = -\alpha \vec k\times \int_0^s\delta \vec t -\beta \int_0^s \delta \vec u.\label{eqn:deltawsol}
    \end{equation}
    \end{subequations}
Lastly, we perform a Chebyshev spectral analysis of the linearized system. Namely, expanding \cref{eqn:deltatwsol,eqn:deltadotu2} as in \cref{nonlinear-numerics}, we obtain a linear dynamical system 
\begin{equation}
    \delta \dot{\mathbb U} = \matr L \delta \mathbb U\label{eqn:linearsystem}
\end{equation}
for the Chebyshev coefficients $\delta\mathbb U$. Note that, since the orthogonality constraint, \cref{eqn:orthogonality}, is stable by \cref{eqn:udott}, we need not consider it in the stability analysis, as coordinates orthogonal to the constraint surface will vanish. The complex eigenvalues of $\matr L$ can be computed numerically; specifically, the system is linearly stable if all the real parts $\omega_i \in \mathbb R^{3\pp{N+1}}$ of these eigenvalues are negative. Here, the system appears to be stable for all values of $\lambda$ and $\omega$ tested. The results are consistent with the dynamics predicted in \cref{base}, which is  dominated by a decay rate of order $\E^{- \beta t} $. %Note that, for a relatively small number of Chebyshev coefficients, negative real parts may be captured (especially when $\lambda$ is large), however, this result does not persist upon increasing $N$, showing that this instability is only numerical. 

\begin{figure}[ht!]
    \centering
    \includegraphics[width=\linewidth]{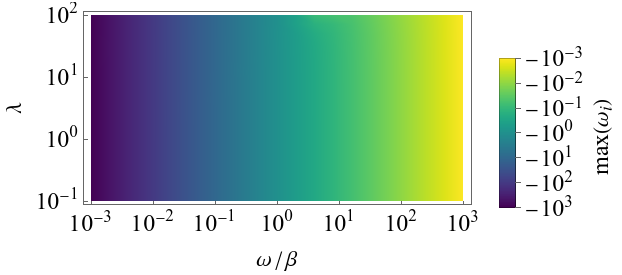}
    \caption{Numerical linear stability analysis. Density plot showing the value of the largest real part $\omega_i$ of the eigenvalues of \cref{eqn:linearsystem} (to generate this plot, the system was re-expressed in terms of the dimensionless time $\omega t$). This shows that the dynamics is dominated by a decay rate of order $\E^{- \beta t} $ as expected from \cref{base}.}
    \label{linear-stability}
\end{figure}

\section{Shoot elongation\label{growth}}
\subsection{General model}
To model growth, we introduce the standard growth multiplier $\gamma \eqdef \linepdiff{s}{s_0}$ which connects the arclength $s_0\in\bb{0,\ell_0}$ in the initial configuration of the shoot, to the arclength $s\in\bb{0,\ell\pp{t}}$ in the current, grown configuration \citep{Goriely2017}. To account for apical dominance, we assume that growth and curvature generation mostly happen within a finite distal section of the stem of length $\delta$. Therefore, we introduce an activation function:
\begin{equation}
    a\pp{s_0, t} = f\pp{\ell\pp{t}-s\pp{s_0,t}},\label{eqn:activation}
\end{equation}
with $f\pp{\sigma} = \E^{-\sigma/\delta}$, modeling the slowing down of growths as we move away from the tip of the shoot, located at $\ell\pp{t} = s\pp{\ell_0,t}$. Accordingly, we assume an exponential growth kinetics given by \citep{Goriely2017}
 \begin{equation}
      \Gamma\eqdef \frac{\dot \gamma}{\gamma} = \Gamma_0 \,  a\pp{s_0, t}, \label{eqn:growth}
    \end{equation}
    which captures a type of growth where all cells in a small portion of the tissue expand and proliferate at the same rate. Similarly, we define the rates of curvature generation $
        A\pp{s_0, t} = \alpha a\pp{s_0,t}$, and $B\pp{s_0, t} = \beta a\pp{s_0,t}$. Note that the model can be easily adapted to include richer apical growth models, e.g. sigmoids \citep{morris1992use}, however, we do not expect any significant qualitative change in the results.

On integrating the standard kinematic relation $ \linepdiff{\dot s}{s}  = \Gamma$ using \cref{eqn:growth,eqn:activation}, we obtain 
\begin{equation}
   \dot s =  c  \E^{-\ell/\delta} \pp{\E^{s/\delta}-1},\label{eqn:dots}
\end{equation}
with $c \eqdef \Gamma_0\delta$ a characteristic speed; and where $\ell $ is governed by
\begin{equation}
   \dot \ell =  c  \left(1-\E^{-\ell/\delta}\right),
\end{equation} as a particular case of \cref{eqn:dots}. Provided the initial condition $\ell\pp{0} = \ell_0 \equiv 1$, the previous equation integrates as
\begin{equation}
    \ell\pp{t} = \delta  \log \pp{(\E^{{1}/{\delta }}-1)\,\E^{\Gamma_0 t}+1}  .\label{eqn:lt}
\end{equation}
Integrating \cref{eqn:dots} with \cref{eqn:lt} then gives
\begin{widetext}
    \begin{equation}
   s\pp{s_0,t} = \delta  \log \bb{\frac{1}{2} -\frac{1}{2}\tanh \left( \frac{\Gamma_0t}{2}+\frac{1}{2\delta}+ \arctan (1-2 \E^{s_0/\delta })-\frac{1}{2}\log \left((\E^{1/\delta }-1) \E^{\Gamma_0 t}+1\right)\right)}.
    \end{equation}
    Thus,
        \begin{equation}
   f\pp{s_0,t}    =  \bb{ \exp\pp{\frac{ct  + 1-s_0 }{\delta }}-\E^{\Gamma_0 t}+1}^{-1},
    \end{equation}
    and
\begin{equation}
\gamma\pp{s_0,t} = \bb{ \left(1-\E^{\Gamma_0 t}\right) \exp\pp{\frac{s_0- c t-1}{\delta }}+1}^{-1}.
\end{equation}
\end{widetext}
In the context of a growing spatial domain, one must differentiate between the material (\textit{Lagrangian}) derivative, denoted with an overdot $\vec{\dot u}$, and the \textit{Eulerian} derivative denoted $\linepdiff{\vec u}{t}$, and such that
\begin{equation}
    \vec{\dot u} = \pdiff{\vec u}{t} + \dot s\pdiff{\vec u}{s}.
\end{equation}
The vectors $\vec u$ and $\vec w$ are defined here in the Eulerian sense, namely such that
\begin{equation}
    \pdiff{\vec t}{s} = \vec u\times \vec t, \quad \pdiff{\vec t}{t} = \vec w\times \vec t,
\end{equation}
with the compatibility condition
\begin{equation}
    \pdiff{\vec u}{t} - \pdiff{\vec w}{s} = \vec w\times \vec u.
\end{equation}
In contrast, the Lagrangian spin vector, $\vec p = \vec w + \dot s \vec u$, is associated with \begin{equation}
    \vec{\dot t}  = \vec p\times \vec t.
\end{equation}
The revised governing equations, including growth, are  then
\begin{subequations}\label{eqn:sys-growth}
\begin{equation}
    \vec t' = \gamma \vec u \times \vec t,
\end{equation}
\begin{equation}
    \vec p' = \gamma   \pp{A \vec t\times \vec k -B \vec u}  ,
\end{equation}
\begin{equation}
      \vec{\dot u} + \vec u\times\vec p + 
 \Gamma\vec u =   \vec p' /\gamma ,
    \end{equation}
\end{subequations}
where $(\ )'$  denotes a derivative with respect to the Lagrangian coordinate $s_0$. The extra term $\Gamma \vec u$ accounts for the passive decrease of curvature due to axial stretch. The presence of the factor $\gamma$ simply results from the chain rule, as we have expressed the system with respect to $s_0$.
\begin{figure}[h!]
    \centering
    \includegraphics[width=.9\linewidth]{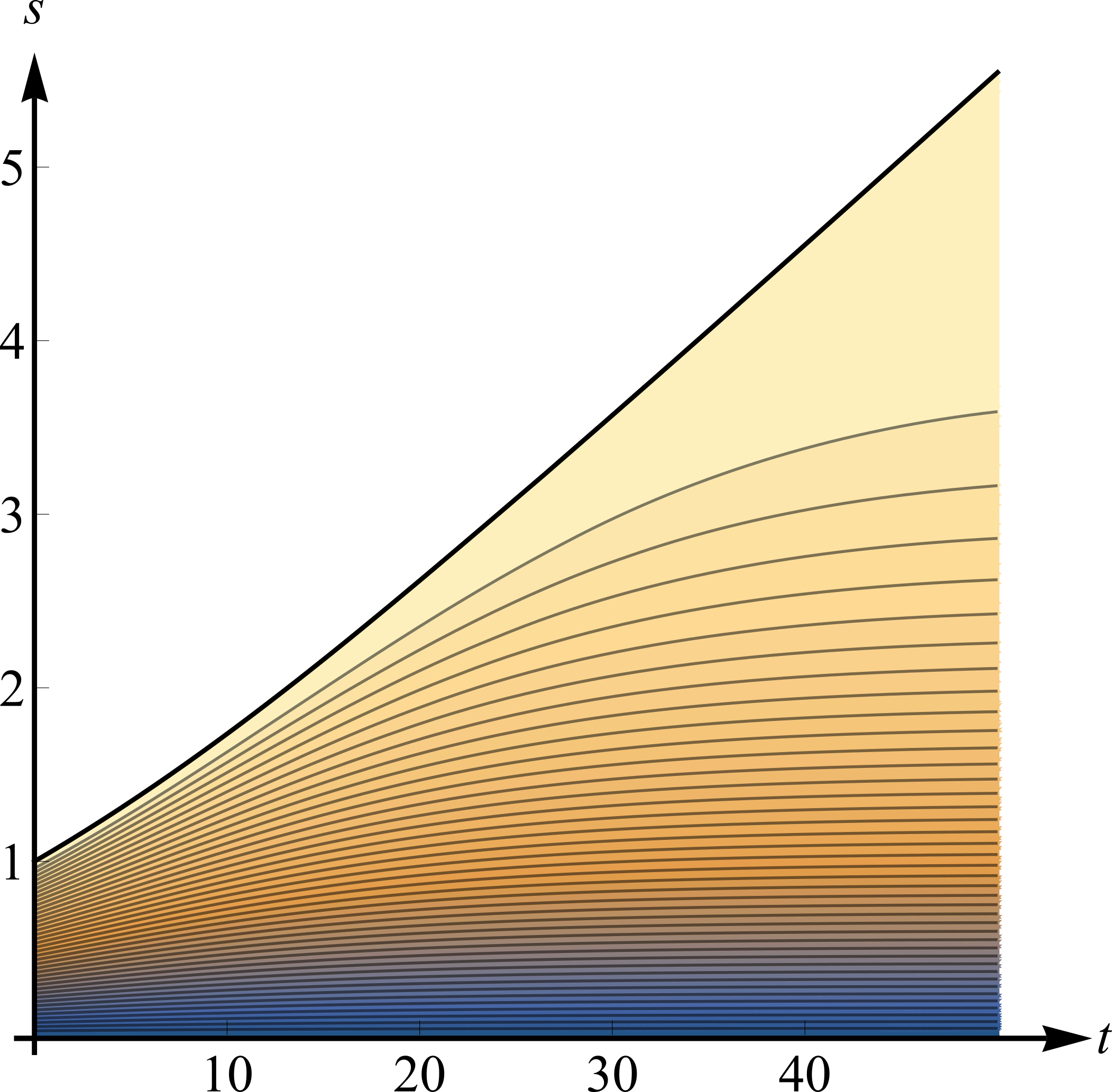}
    \caption{Kymograph showing the apical growth field. Lines show the trajectories of the material points with initial arclength emphasized by colors.}
    \label{fig:growth}
\end{figure}

\subsection{Solitary waves}
To derive the shape of self-similar, traveling-front solutions we introduce the co-moving coordinate
 $\sigma\eqdef  \ell - s$, measuring the arclength from the apex, with the base located at $\sigma = \ell  \rightarrow \infty$. Setting $\linepdiff{\vec u}{t}=\vec 0$, \cref{eqn:sys-growth} becomes upon this change of coordinate:
\begin{subequations}\label{eqn:tw}
\begin{equation}
    \pdiff{ \vec t}{\sigma} = \vec t\times \vec u ,
\end{equation}
\begin{equation}
    \pdiff{\vec p}{\sigma} =  f\pp{\sigma}\pp{\alpha \vec k\times\vec t +\beta \vec u},
\end{equation}
\begin{equation}
      c f\pp{\sigma}\pdiff{\vec u}{\sigma} + \pdiff{\vec p}{\sigma} = \vec p\times\vec u - \Gamma \vec u,\label{eqn:solitarywaveu}
    \end{equation}
\end{subequations}
%with $\dot s = \dot\sigma - c = c f\pp{\sigma}$, and
with the conditions $\displaystyle\lim_{\sigma\rightarrow\infty}\vec t=\vec i$, $\displaystyle\lim_{\sigma\rightarrow\infty}\vec p=\omega \vec i$ and $\displaystyle\lim_{\sigma\rightarrow\infty}\vec u=\vec 0$. In practice, the system can be integrated for $\sigma \in \bb{0,\Sigma}$ with $\Sigma \gg \delta $, and with boundary conditions expressed at $\Sigma$. There is however a removable singularity at $\sigma \rightarrow \infty$, as $f\pp{\sigma}$ is transcendentally small, which causes numerical difficulties in \cref{eqn:solitarywaveu}. To alleviate this issue, we consider perturbed boundary conditions of the form
$
    \vec t\pp{\Sigma} = \vec i + \boldsymbol{\epsilon}_t\pp{\Sigma}
$, $ \vec p\pp{\Sigma} = \omega \vec i + \boldsymbol{\epsilon}_p\pp{\Sigma},
$ and $    \vec u\pp{\Sigma} = \boldsymbol{\epsilon}_u\pp{\Sigma}$, 
where $\boldsymbol{\epsilon}_t$, $\boldsymbol{\epsilon}_p$ and $\boldsymbol{\epsilon}_u$ denote small perturbations from the boundary conditions at $\sigma = \infty$. Expanding \cref{eqn:tw} and keeping only the higher order non-zero terms allows to solve for $\boldsymbol{\epsilon}_t$, $\boldsymbol{\epsilon}_p$ and $\boldsymbol{\epsilon}_u$, in order to express the perturbed boundary values [\cref{fig:fig5}(b) is obtained with $\Sigma \approx 5 \delta $].

\section{Code availability}

All numerical methods were implemented in \textit{Wolfram Mathematica 13.0}. Source code will be made publicly available upon acceptance of the manuscript for publication.

\section{Supplementary files}

\paragraph{Movie 1. --} An example rotating shoot converging towards equilibrium [parameters as in \cref{fig:fig4}(d), with $\beta=\omega/5$].
\paragraph{Movie 2. --} In the absence of autotropism, a shoot will orbit around a caulinoid [parameters as in \cref{fig:fig4}(d), with $\beta=0$].
\paragraph{Movie 3. --} Example traveling solution [parameters as in \cref{fig:fig5}(a), left-hand side simulation].
\paragraph{Movie 4. --} Uniform growth along a caulinoid [$\alpha=5\omega$, $\beta=\omega$, $\Gamma_0 =\omega/10$, $\delta=100$].

\bibliographystyle{apsrev4-2}
\bibliography{biblio}

% Citations in Supp Mat
% \nocite{Hale2015,press2007numerical}

\end{document}